\documentclass[12pt,english]{IEEEtran}
\usepackage[T1]{fontenc}
\usepackage[latin9]{inputenc}
\usepackage{geometry}
\usepackage{float}
\usepackage{units}
\usepackage{enumitem}
\usepackage{amsmath}
\usepackage{graphicx}

\makeatletter
\newlength{\lyxlabelwidth}      
\newcommand{\lyxaddress}[1]{
	\par {\raggedright #1
	\vspace{1.4em}
	\noindent\par}
}
	\newenvironment{elabeling}[2][]%
	{\settowidth{\lyxlabelwidth}{#2}
		\begin{description}[font=\normalfont,style=sameline,
			leftmargin=\lyxlabelwidth,#1]}
	{\end{description}}

\usepackage{gensymb}

\makeatother

\usepackage{babel}
\begin{document}
\title{Recycling rejected silicon wafers and dies for high grade PV cells}
\author{G. Golan{*}\thanks{{*}Corresponding author, email: gadygo@ariel.ac.il},
M. Azoulay and G. Orr}
\maketitle

\lyxaddress{Ariel University, Ariel 40700, Israel}
\begin{abstract}
The recent return of the US to the Paris Climate Accord, massive increase
in solar panel production and energy storage solutions has resulted
in pressure on supply for solar cell materials and recycling of panels
installed in the 90's and beginning of the 2000's which have reached
their end of life. In this work we focus on recycling silicon wafers
and dies by stripping previous structures from the die using potent
acids after which its base material is characterized and binned. We
demonstrate the process for silicon p-type substrates where n-type
doping is attained by using a simple solution of phosphoric acid,
which is diffused into the substrate using a furnace thus creating
a PN junction. In case the substrate is n-type it could be replaced
by boric acid. This is followed by deposition of a conductive antireflective
coating, bus bars and rear wafer metal coating. The initial demonstrated
laboratory results indicate the feasibility of recycling wafers using
simple low cost standard industrial methods. 
\end{abstract}


\section{Introduction}

Renewable energy sources for electrical energy generation has become
an increasing concern of most countries due to the environmental impact
associated with the use of fossil fuels. In addition, reducing a countries
dependence on fossil fuel further increases its self reliance and
improves a nations balance of trade, thus boosting its economy. Some
countries have initiated programs increasing economic incentives to
companies and individual households for developing, improving efficiencies
or increasing capacity of renewal energy sources. In a recent report
\cite{van2019new} the International Renewable Energy Agency estimates
the current cost of PV electrical generation at optimal locations,
is 30 USD/MWh and predicts that by 2022 it will reach prices that
are comparable to the lower end of fossil fuel costs. The various
steps of manufacturing solar panels from raw materials is energy intensive
and includes the use of large amounts of water and toxic chemicals.
Therefore, expecting the rapid growth of silicon use for solar energy
generation since the late 90's of the previous century, research was
conducted for initiating environmentally benign solar cell manufacturing
\cite{tsuo1998environmentally}. The predicted increase in solar panel
production and installations \cite{van2019new} and the 25 year life
expectancy of a solar panel has initiated various End of Life (EOL)
management programs \cite{corcelli2018sustainable,padoan2019recycling,shin2017method,weckend2016end}.
The rapid growth of solar energy generation having a capacity of 222
GW in 2015 and expected to rise up to 4,500GW by 2050 \cite{weckend2016end}
will result in an increase of PV panel waste. The dominant technology
which has developed rapidly since the 70's of the previous century
is the photovoltaic solar cell technology. Since the mid 90's of the
previous century, due to the leading industrialized countries governmental
policies, silicon solar photovoltaic energy generation has become
commercially viable. Most silicon solar cells are based on single
crystal silicon (monocrystals) grown by the Czochralski technique
and polycrystalline silicon casting. While there are more efficient
materials for photovoltaic energy generation than single crystal silicon,
they are not commercially viable for large area application. In 2015,
crystalline silicon had approximately 93\% of the market share of
solar modules, with 24\% going to the monocrystals and 69\% going
to the polycrystalline solar cells \cite{placzek2017top}. The larger
market share of the polycrystalline solar cells is due to the significantly
lower production costs, but this comes at a price of lower efficiency
compared to the single crystal silicon. For most uses, where space
or weight are not a concern the polycrystalline solar cells are appropriate.
In places where space and weight are restricted, single crystal silicon,
due to their higher efficiency are better suited. When dealing with
monocrystal based photovoltaic cells, higher quality crystals result
in higher conversion efficiency, but it comes at an economical price.
Integrated circuit grade silicon is probably the most efficient base
material for fabricating silicon solar panels due to its low defect
density, but the lengths and costs for obtaining it, is prohibitive
for its wide application in terrestrial energy generation, where the
capital investment per Watt is the key driving force. In order to
reduce costs solar grade silicon monocrystals with a much higher level
of defects and impurities are being introduced \cite{delannoy2012purification}.
Another concern which needs to be addressed is recycling of the panels
when they reach their end of life (EOL). In this respect, rare metals
and the silicon cells retain their value and should be salvaged using
various techniques \cite{shin2017method,padoan2019recycling}. One
untapped resource of high quality near defect free silicon, are rejected
dies and wafers. With an estimated 90\% yield during full production
\cite{van2000microchip} we are left with 10\% rejected dies. Currently,
major fabs are producing integrated circuits with wafer diameters
larger than $400\,mm$. This provides an abundant supply of silicon
for producing high quality solar panels. In this article we will present
a relatively environmentally friendly method for recycling silicon
from the rejected wafers. In a following article, we will present
a method for assembling an efficient solar panel from the salvaged
wafers.

\section{Materials preparation}

This work based on experiments conducted by our groups, suggests a
method for recovering rejected silicon dies, wafers and residual silicon
which resides on the rims of the processed wafers. In order to keep
the text clear, we shall generally refer to all the recovered silicon
as dies. The process results in PN photovoltaic structures composed
of $In_{2}O_{5}Sn/n-Si/p-Si$ $ITO,n-type\,silicon,\,p-type\,silicon$. 

\subsection{Stripping}

The dies were etched in $HF$ after which the conductivity and type
were measured. This assists us with defining the type of base material
to deposit prior to diffusion (Phosphorous/Arsenic or Boron/Gallium).
Similar die types are binned together. In this article we shall describe
the process for a p-type substrate. Standard cleaning recipe was applied
to the dies with an optional oxide strip using $HF$ between the first
step (SC-1) and the second step (SC-2). When using those steps, the
rare metals can be recovered using the processes described in \cite{padoan2019recycling}. 

\subsection{Donor/Acceptor diffusion}

The samples were spin coated using a solution of 1:1 ratio of phosphoric
acid and ethanol ($H_{3}PO_{4}:C_{2}H_{5}OH$). If the substrate is
n-type one can replace the phosphoric acid with boric acid ($H_{3}BO_{3}$).
The samples were place in a furnace with an ambient atmosphere pre-baking
them at a temperature of $150\degree C$ for 10 minutes drying the
newly applied coating. The temperature was raised to a temperature
of $900^{\degree}C$ for 6 hours slowly cooling it to a level of \textbf{$600\degree C$},
this procedure allows for propper diffusion while reducing destructive
glass phases. Dopant concentration was evaluated using ICP-OES \cite{Rietig2017}
which can trace down to approximately $N_{d}\sim10^{14}\,[cm^{-3}]$.

\subsection{Transparent Conductive Oxide coating}

Although much work was conducted in the past 20 years, on the topic
of environmentally sustainable, indium free Transparent Conducting
Oxides (TCOs), for example \cite{Guilln2011TCOmetalTCOSF,Fortunato2007TransparentCO},
in this work we used ITO with the standard application methods. 

\begin{figure}[h]
\centering{}\includegraphics[scale=0.45]{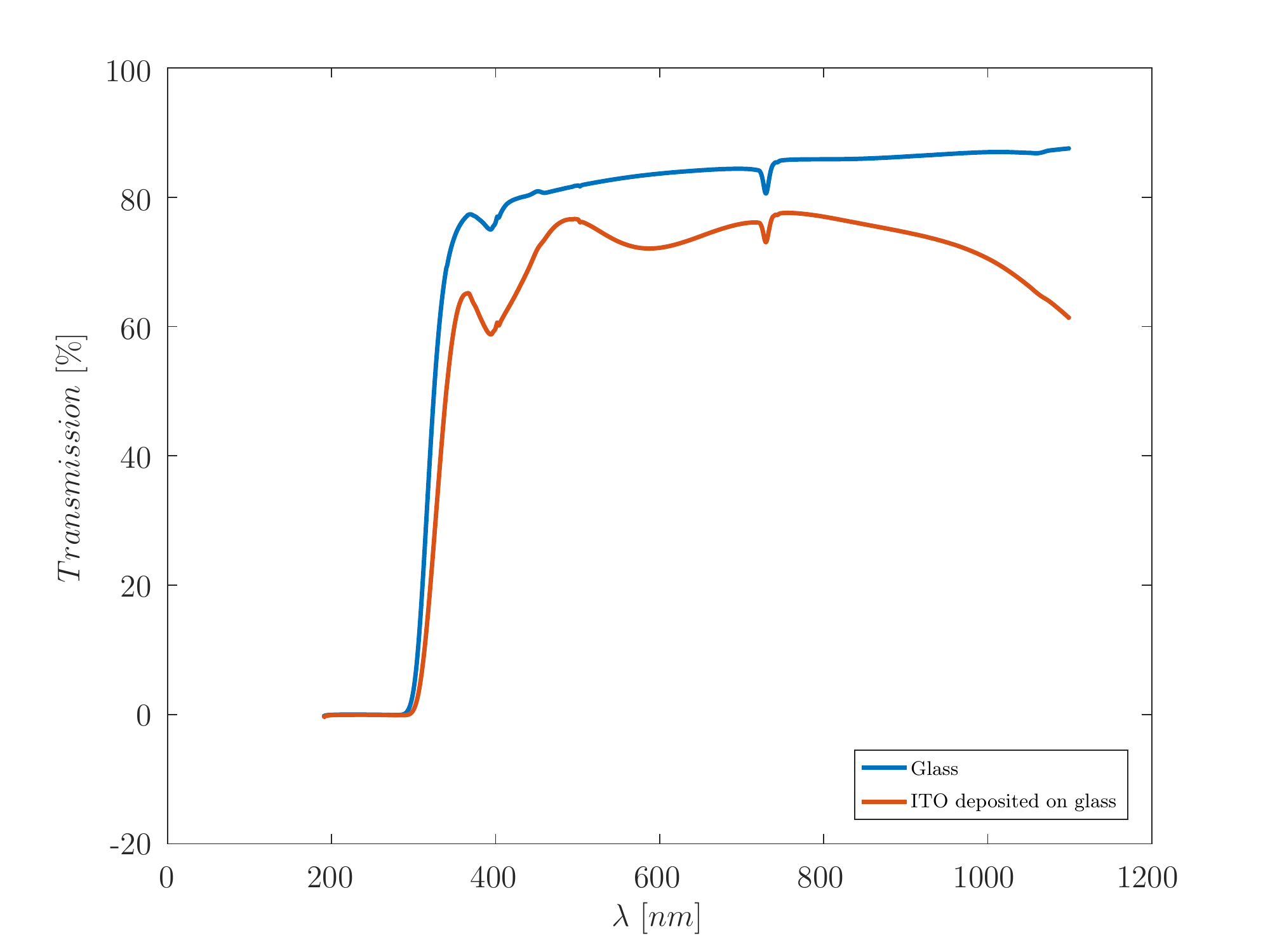}\caption{\label{fig:Optical-transmission-of}Optical transmission of the ITO
on a glass substrate compared to the transmission of the glass substrate.}
\end{figure}

Transparent conductive films of $In_{2}O_{5}Sn$ were deposited by
RF sputtering using an ITO ceramic ($90\,wt\,\%\,In_{2}O_{3},\,10\,wt\,\%\,SnO_{2}$).
Substrates were cleaned using ultra pure deionized water isopropyl
alcohol and acetone in an ultrasonic cleaner bath for 20 minutes.
This was followed by drying using a flow of nitrogen gas. The RF sputtering
chamber was purged with Ar and a working pressure of $100\,mTorr$
was maintained. Surface oxidation was removed in the pre-sputtering
stage. A film of approximately 250nm was deposited on the surface
of the materials with the procedure tested using microscope slides
prior to the actual deposition. The deposited film was annealed in
a tube furnace in an oxygen saturated environment supported by a 0.5
sccm $O_{2}$ gas flow at a temperature of $400\degree C$. Optical
transparency of the $In_{2}O_{5}Sn$ films was measured in the UV/VIS
using a Jasco V-730. Additionally, it was used in reflective mode
to measure the film thickness \cite{Arndt:84} using the SLM-907 specular
reflectance accessory. Figure \ref{fig:Optical-transmission-of} illustrates
the transmission of the glass substrate and the ITO deposition on
the glass substrate.

\begin{figure}[H]
\centering{}\includegraphics[scale=0.45]{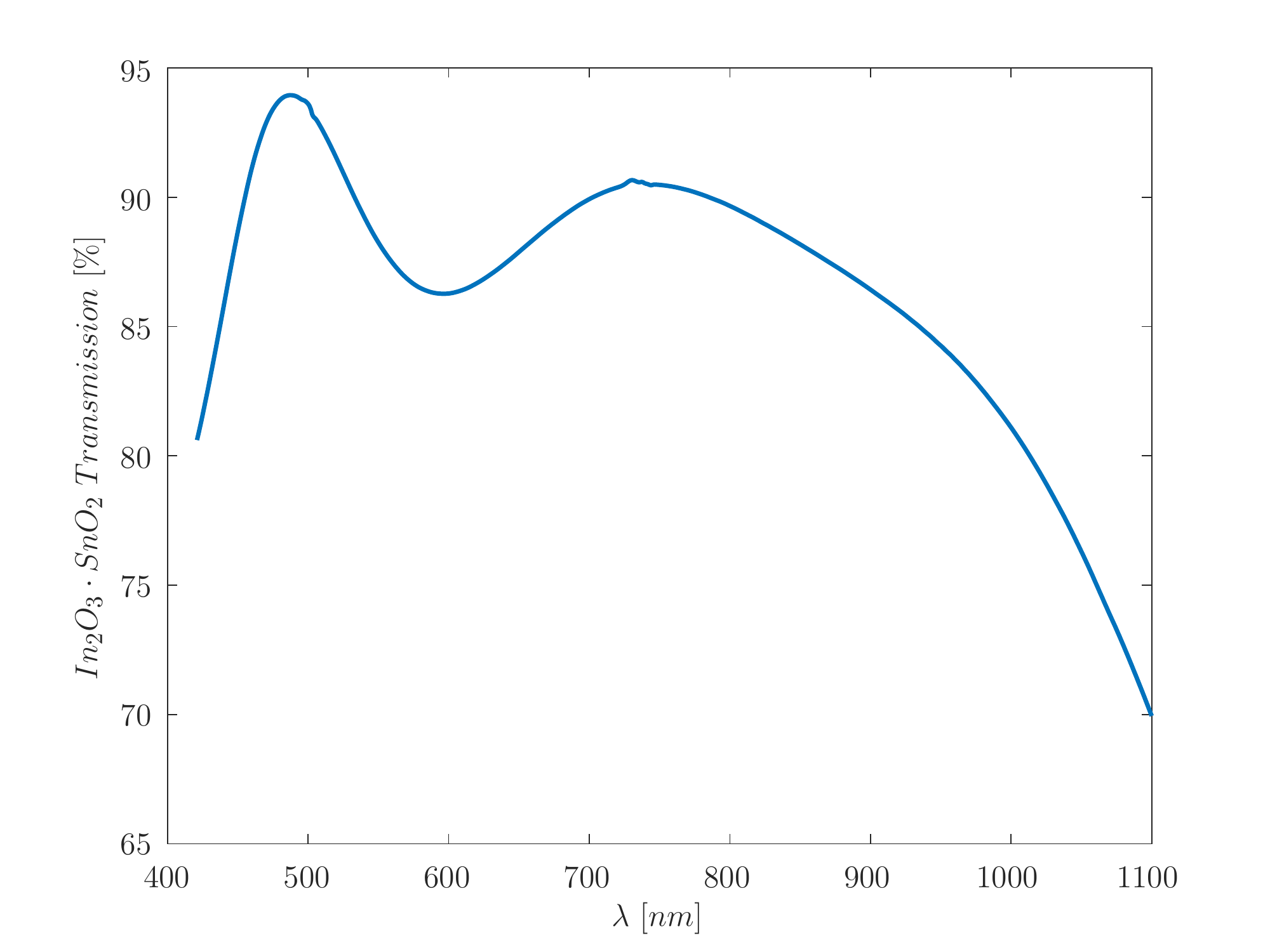}\caption{\label{fig:Optical-transmission-excluding-glass}Optical transmission
of the ITO excluding the glass substrate}
\end{figure}

The optical transmission of the ITO excluding the glass substrate
in the 420-1100nm range is given in figure \ref{fig:Optical-transmission-excluding-glass}.

As can be seen in figure \ref{fig:Optical-transmission-excluding-glass},
the optical transmission at 500nm is 93.6\% which compares favorably
with published data. Inkjet printing of the collector bus bar and
finger aluminum electrodes were tested with inconclusive results.
As a result we opted for the traditional method of applying aluminum
electrodes by thermal evaporation deposition.

\begin{figure}[H]
\centering{}\includegraphics[scale=0.45]{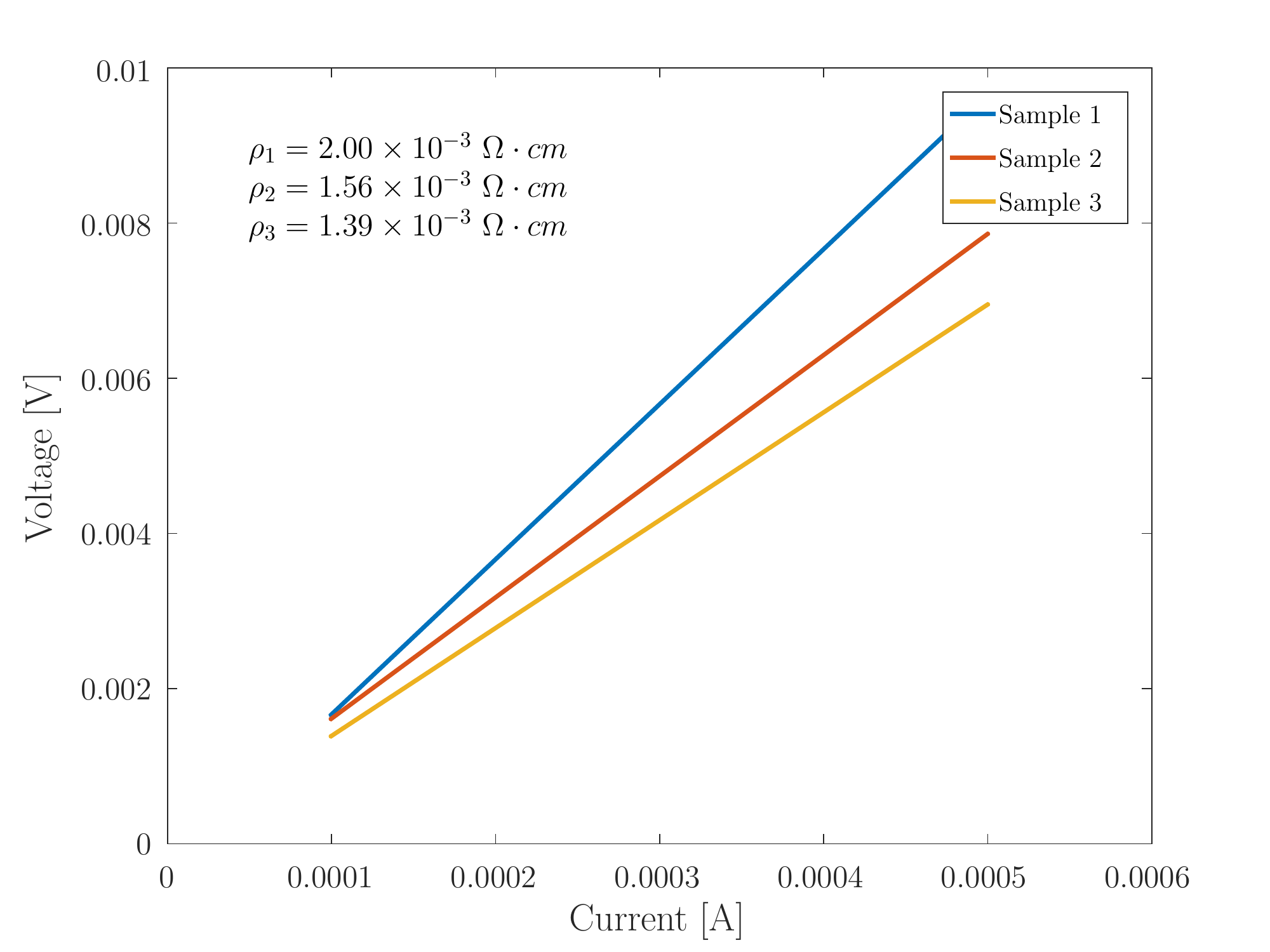}\caption{\label{fig:Resistivity-of-of}Resistivity of the ITO deposited layer
of three different batches as extracted from a four point measurement}
\end{figure}

A four-point measurement of the $In_{2}O_{5}Sn$ deposited layer resistance
was conducted before and after the diffusion validating the resistivity
of the layer. Figure \ref{fig:Resistivity-of-of} illustrates the
resistivity obtained using three different samples from three different
batches.

The resistivity results correspond with published data \cite{kim2009preparation}.
A hot probe experiment identified the type of semiconductor layer
after the diffusion process a detailed account of the measurement
can be found in \cite{van2004principles}. During the various processes
the surface structure was studied and documented using a metallurgical
microscope. 

\section*{Experimental Results}

\begin{figure}[H]
\centering{}\includegraphics[scale=0.6]{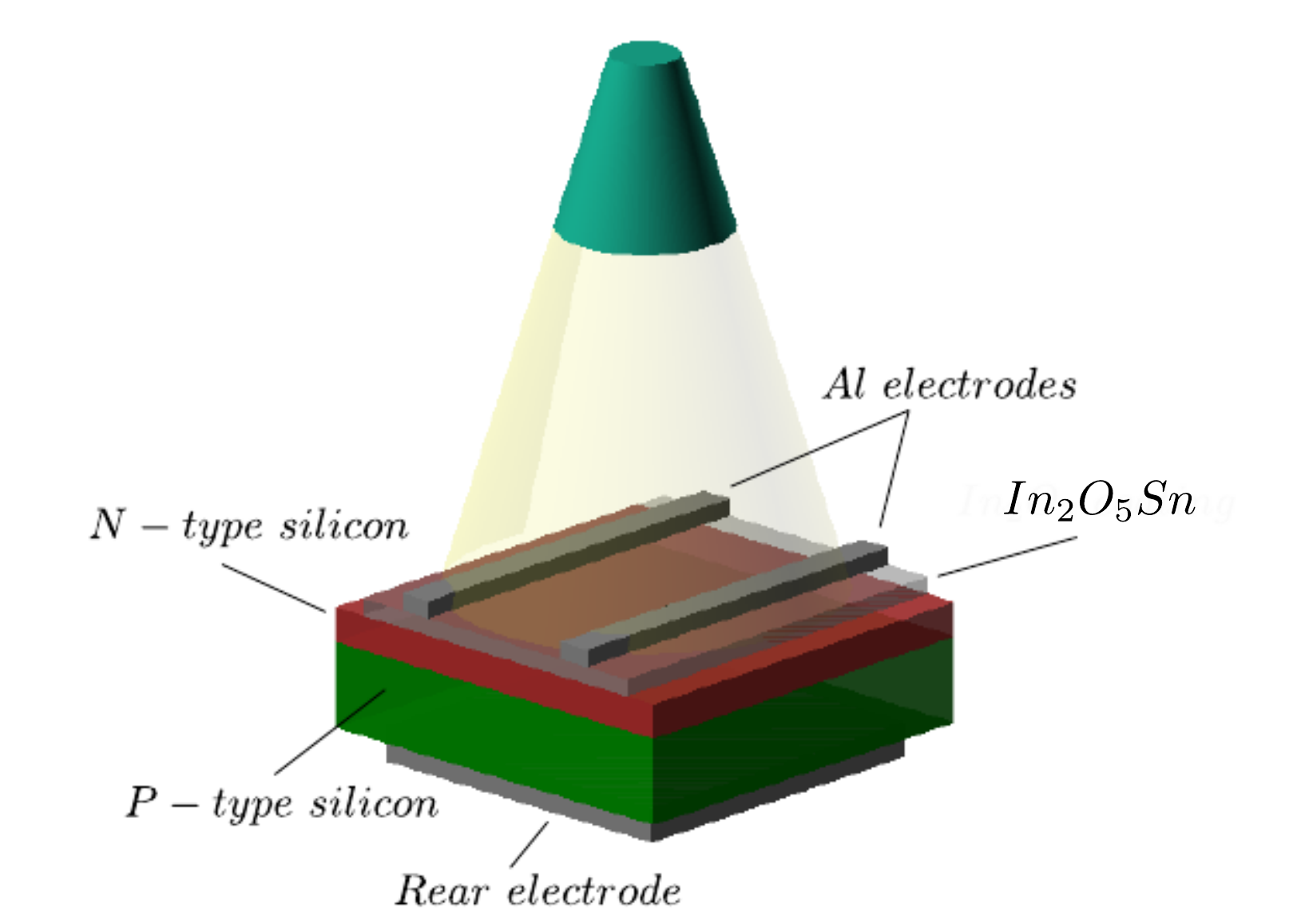}\caption{\label{fig:Photovoltaic-cell-structure}Photovoltaic cell structure}
\end{figure}

A standard photovoltaic structure with a thin emitter fabricated at
the top surface using a p-type substrate is demonstrated in figure
\ref{fig:Photovoltaic-cell-structure}.

The wavelength dependent short circuit current is approximated by
\begin{equation}
J_{SC}(\lambda)=\frac{q\phi(\lambda)(1-R_{f})\alpha L_{p}}{\alpha^{2}L_{p}^{2}-1}\varPhi_{e}\label{eq:Short_circuit_current}
\end{equation}
with $\Phi_{e}$ presenting an emitter factor which depends on doping
levels and depth of doping:\begingroup
\small
\begin{align}
\Phi_{e} & =\frac{\frac{S_{p}L_{p}}{D_{p}}+\alpha L_{p}-e^{-\alpha H_{e}}\left(\frac{S_{p}L_{p}}{D_{p}}\cdot\cosh(\nicefrac{h_{e}}{L_{p}})+\sinh(\nicefrac{h_{e}}{L_{p}})\right)}{\frac{S_{p}L_{p}}{D_{p}}\sinh(\nicefrac{h_{e}}{L_{p}})+\cosh(\nicefrac{h_{e}}{L_{p}})}\label{eq:emitter_geometry_factor}\\
 & -\alpha L_{p}e^{-\alpha H_{e}}\}\nonumber 
\end{align}

\endgroup

where:
\begin{elabeling}{00.00.0000}
\item [{$\phi(\lambda)$}] photon flux $\left[\nicefrac{\#photons}{sec\cdot m^{2}}\right]$
\item [{$R_{f}$}] reflectance of the active surface of the cell
\item [{$\alpha$}] absorption coefficient of the semiconducting material
$[cm^{-1}]$
\item [{$L_{p}$}] diffusion length of the minor charge carriers in the
emitter $L_{p}=\sqrt{D_{p}\cdot\tau_{p}}$ $[cm]$
\item [{$S_{p}$}] surface recombination velocity for minor charge carriers
$[cm/s]$
\item [{$h_{e}$}] depth of the emitter $[cm]$
\item [{$D_{p}$}] diffusion coefficient of minor charge carriers in the
emitter $[cm^{2}/s]$
\end{elabeling}
$J_{ph}=q\cdot\phi(\lambda)$ the photon flux current is referred
to as the photon flux which is the current given that every photon
generates an electron hole pair. In order to get a grasp of the behavior
of the short circuit current (equation \ref{eq:Short_circuit_current})
one has to observe that the emitter factor (equation \ref{eq:emitter_geometry_factor})
has the form
\[
\Phi_{e}\approx C-Be^{-t}\underset{t\rightarrow\infty}{\longrightarrow}C
\]
In which C is a constant depending on the minor charge carrier attributes
$(L_{p},D_{p})$ which are donor density dependent, the depth $h_{e}$
and the surface recombination velocity $S_{p}$. The surface recombination
velocity depends on both the surface defects and doping levels. For
a passivated surface the doping level is $N_{s}\approx10^{18}-10^{19}$.
The surface recombination velocity is given by \cite{morales1991theoretical,stem2001studies}
\[
S_{p}=10^{-16}N_{s}
\]

\begin{figure}[H]
\centering{}\includegraphics[scale=0.45]{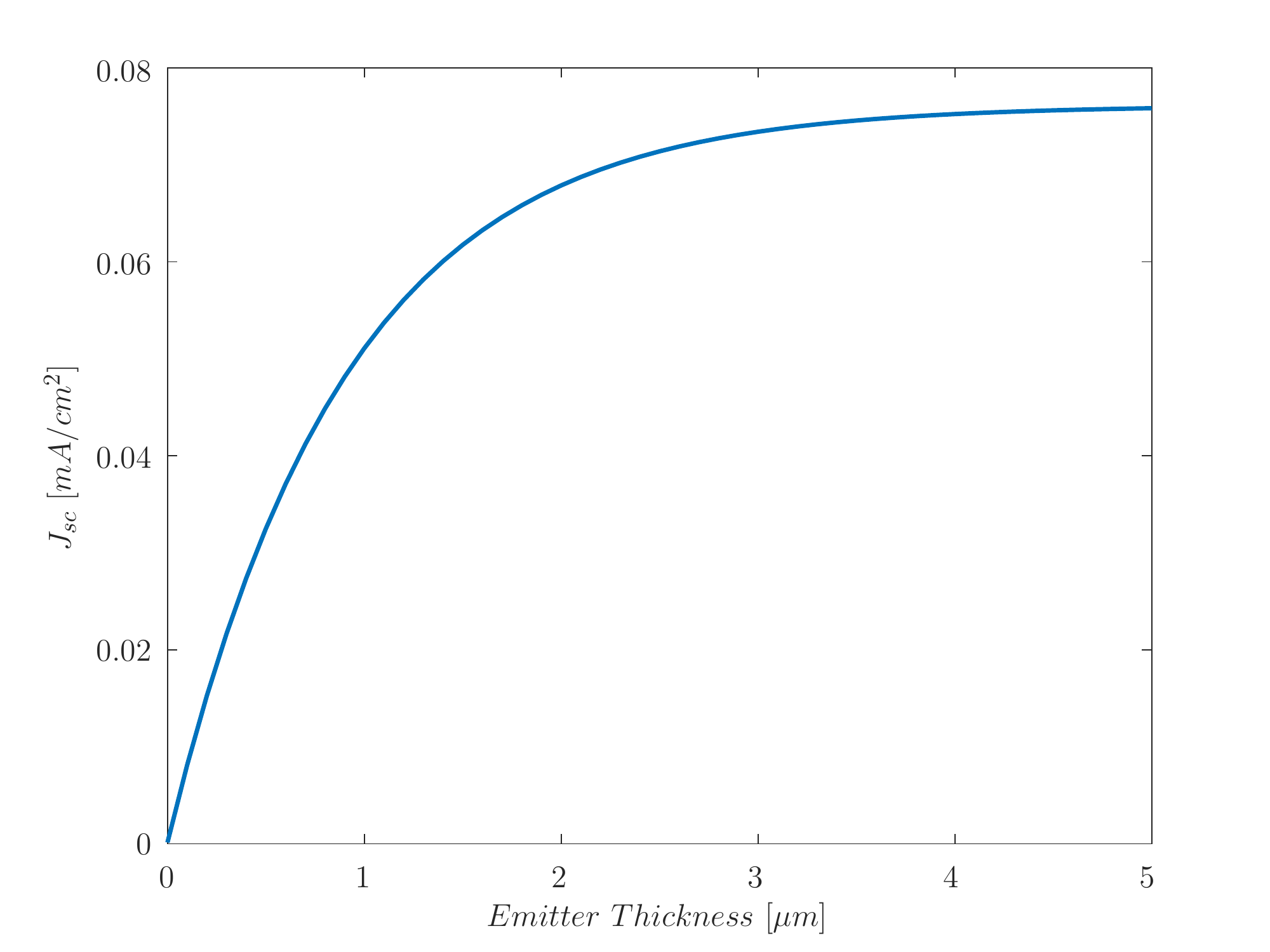}\caption{\label{fig:Short-circuit-current}Short circuit current as a function
of the depth of the emitter for: $\lambda=500nm;R_{f}(\lambda)=0.05\,\alpha(\lambda)=1.11\times10^{4};\,S_{p}=200\,[cm/s];D_{p}=13\,[cm^{2}/s];\,\tau_{p}=10\mu s;J_{ph}(\lambda)=0.08\,[mA/cm^{2}]$}
\end{figure}

Figure \ref{fig:Short-circuit-current} shows the calculated short
circuit current as a function of the depth of the emitter.

For the above parameters it can be seen that the optimal depth of
the emitter is approximately $3\mu m$. The current is wavelength
dependent on three accounts. Two external ones which are the photon
flux, and the reflection $R_{f}$ and an internal one being the absorption
coefficient $\alpha$. 

\subsection{Reflectance}

\begin{figure}[H]
\begin{centering}
\includegraphics[scale=0.45]{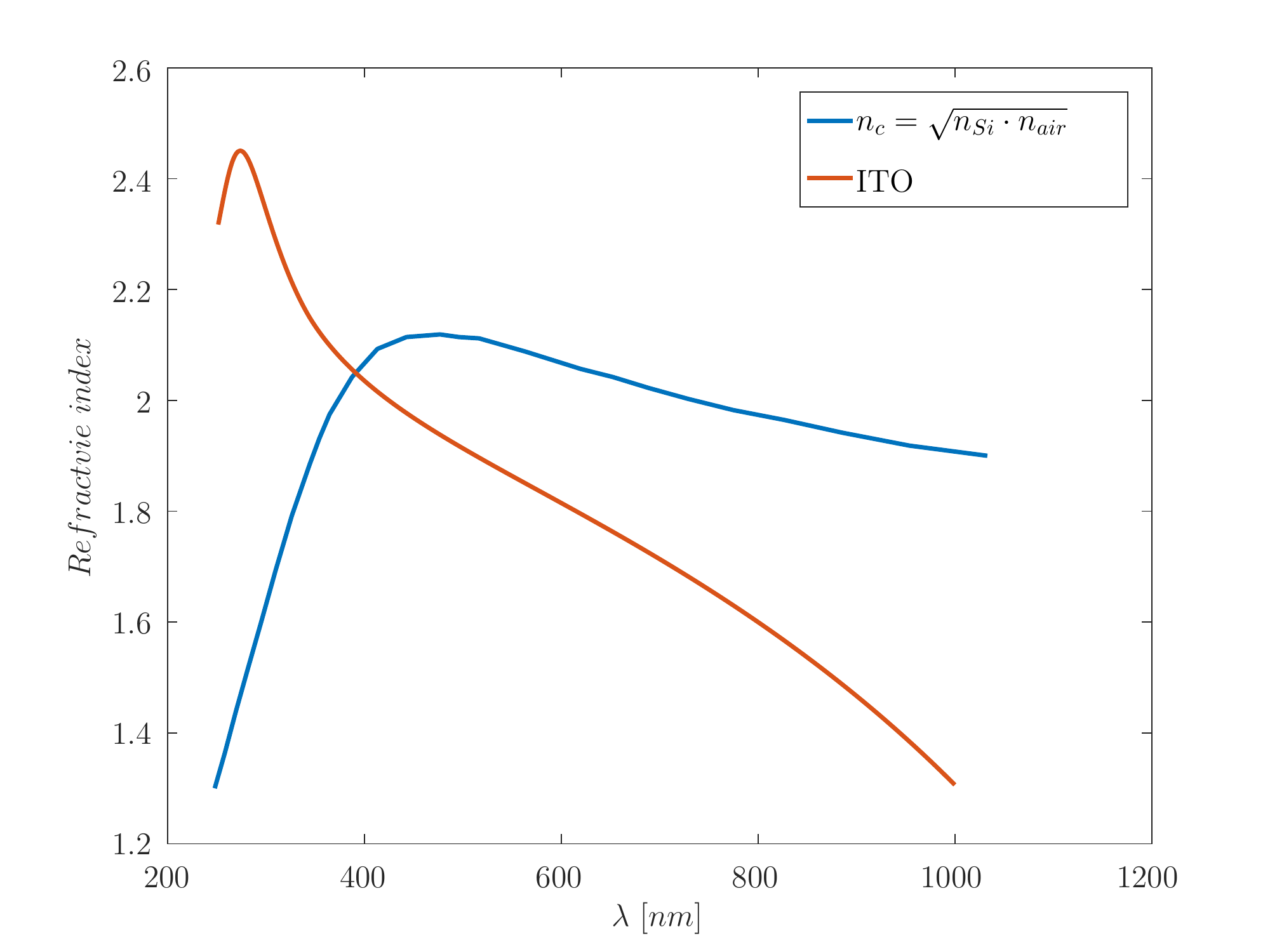}
\par\end{centering}
\caption{\label{fig:Comparison-between-the}Comparison between the index of
refraction of ITO and the square root of $n_{Si}$ and $n_{air}$}
\end{figure}

$R_{f}$ is a significant parameter as it has a considerable influence
on the efficiency of the PV cell as is seen in equation \ref{eq:Short_circuit_current}.
On bare silicon $R_{f}>30\%$ reducing the efficiency considerably.
In order to decrease the reflection loss in the visible and NIR range
an anti-reflective coating is applied. The applied coating needs to
shift the phase of the reflected light by half a wavelength, thus
canceling each other out. This happens when the surface of the silicon
and the applied coating are at a distance of $\lambda/4$, or 
\[
n_{c}d=\lambda/4
\]

a further improvement (minimization of reflection losses) is obtained
when the coating is the mean of the refractive indices ie.
\begin{align*}
n_{c}(\lambda) & =\sqrt{n_{air}\cdot n_{Si}}\\
n_{c}(500nm) & \approx2.11
\end{align*}
one such material that is both transparent and which provides an appropriate
solution over most of the visible range is ITO. Figure \ref{fig:Comparison-between-the}
displays the refractive index and the square root of the refractive
index of the silicon compared to the refractive index of ITO. 

Our peak wavelength is about $500$nm and according to the graph $n_{ITO}(500)\approx2$
thus a deposition thickness of $125nm$ is appropriate for our purpose.
As was seen in figure \ref{fig:Resistivity-of-of}, the low resistivity
of the ITO film on our samples accommodates the conduction of charge
to the fingers and bus bars. 

\subsection{Post diffusion}

\begin{figure}[H]
\centering{}\includegraphics[scale=0.45]{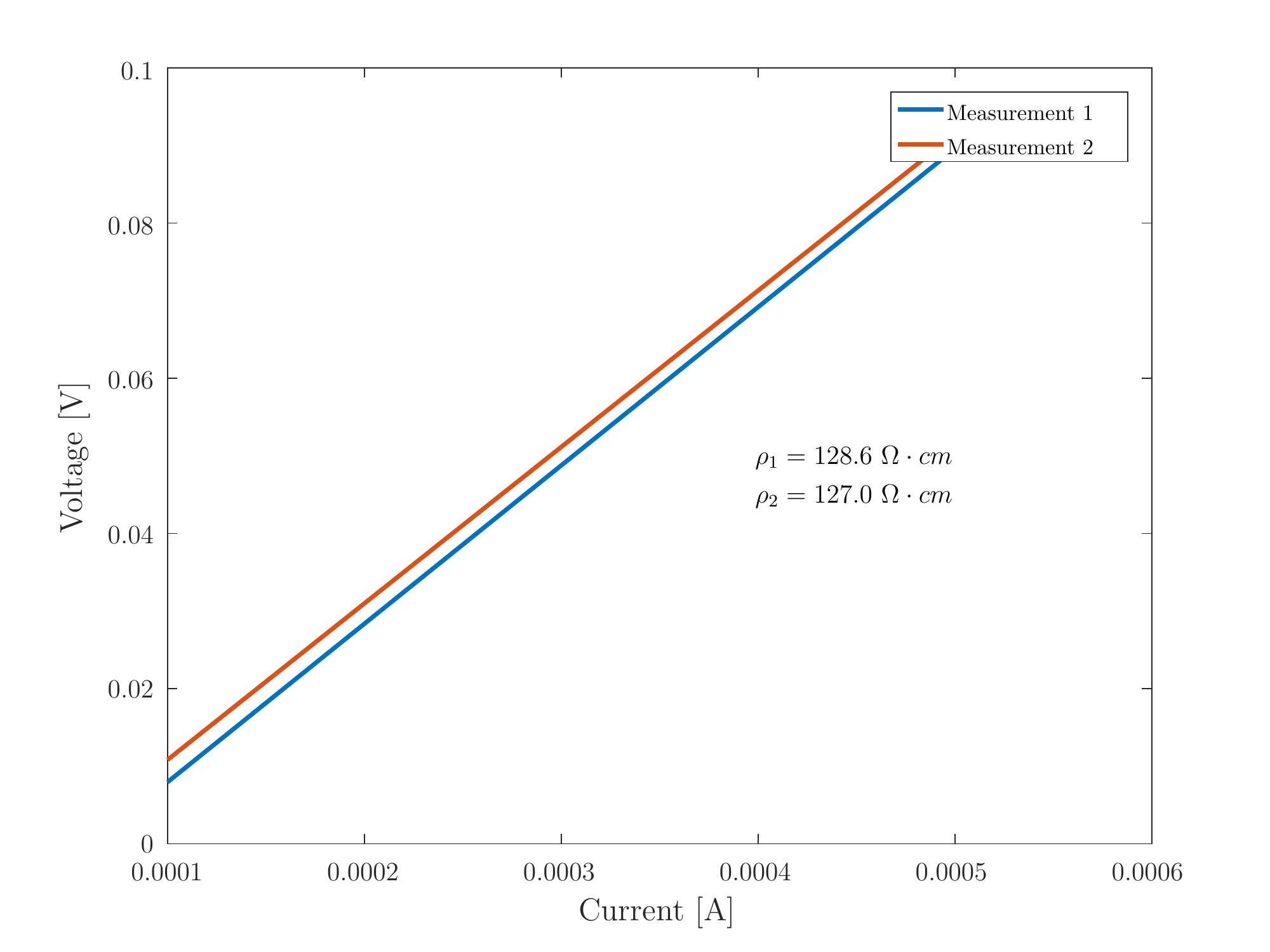}\caption{\label{fig:A-four-point}A four point measurement of the doped (emitter)
layer}
\end{figure}

The describe process's weakness is that, to date, it is not a completely
controlled process. While during the classic method the flow of the
dopant gas is controlled, some of the spin coated material evaporates
and we need to resort to ICP-OES measurements in order to estimate
dopant concentration. The resistivity of the doped layer was measured
using a four point measurement as shown in figure \ref{fig:A-four-point}

\begin{figure}[H]
\begin{centering}
\includegraphics[scale=0.45]{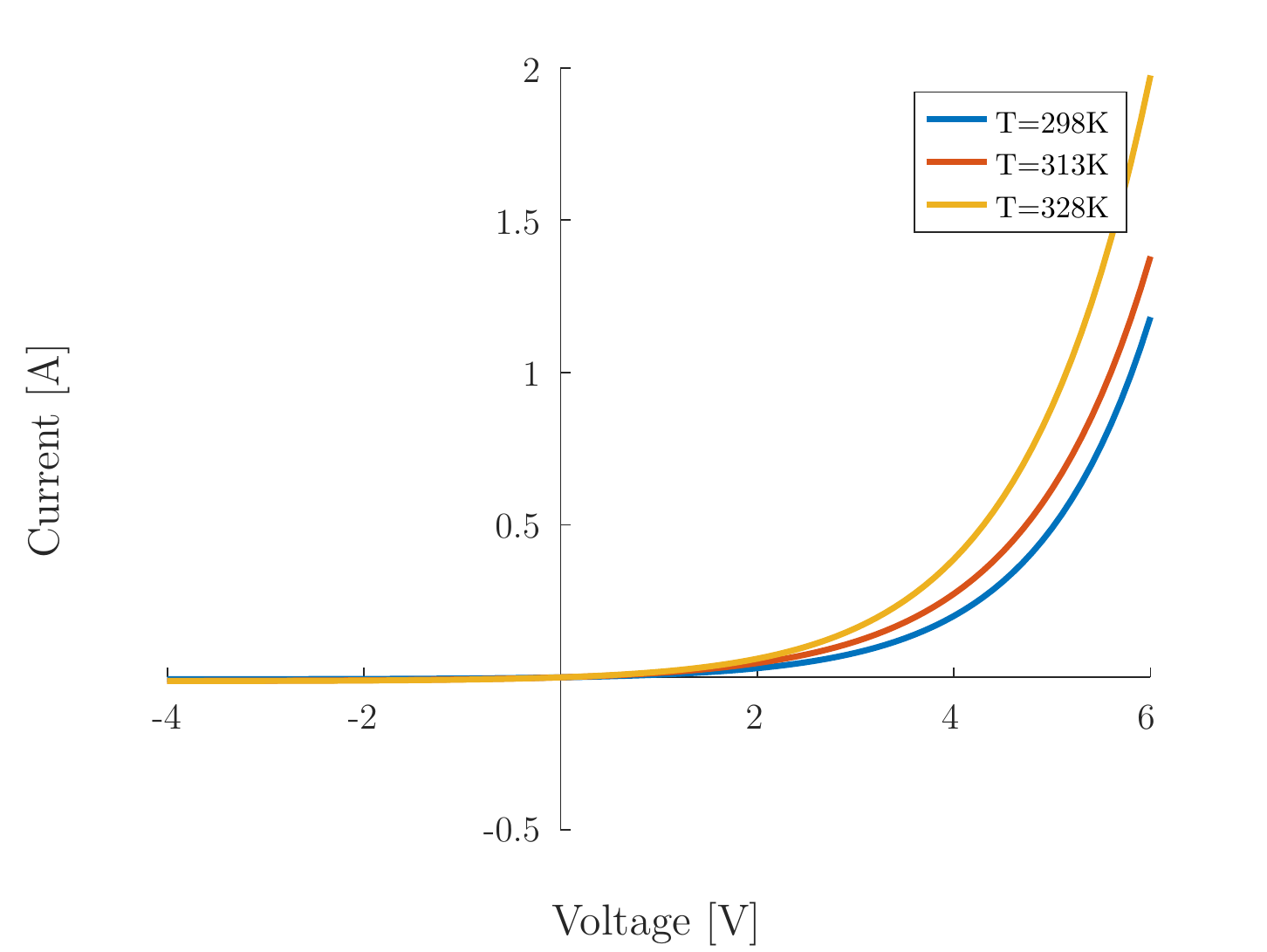}
\par\end{centering}
\caption{\label{fig:Device-I-V-characteristic}Device I-V characteristic at
several temperatures}
\end{figure}

The resistivity result of $\sim127\,\Omega\cdot cm$ corresponds with
a dopant density of $4\times10^{14}\,cm^{-3}$ \cite{thurber1980resistivity}
which was confirmed by the ICP-OES measurement which is approximately
$5\times10^{14}\,cm^{-3}$ (close to the detection limit of the phosphorus).
Following the resistivity characterization the I-V characteristic
was measured at different temperatures. Figure \ref{fig:Device-I-V-characteristic}
indicates that the material displays a regular PN junction behavior.

The calculated potential of the junction is 0.5 V while the actual
open circuit voltage was found to be lower than 0.1 V. This is due
to the high losses in this device which is in its preliminary research
stages. 

\begin{figure}[H]
\centering{}\includegraphics[scale=0.45]{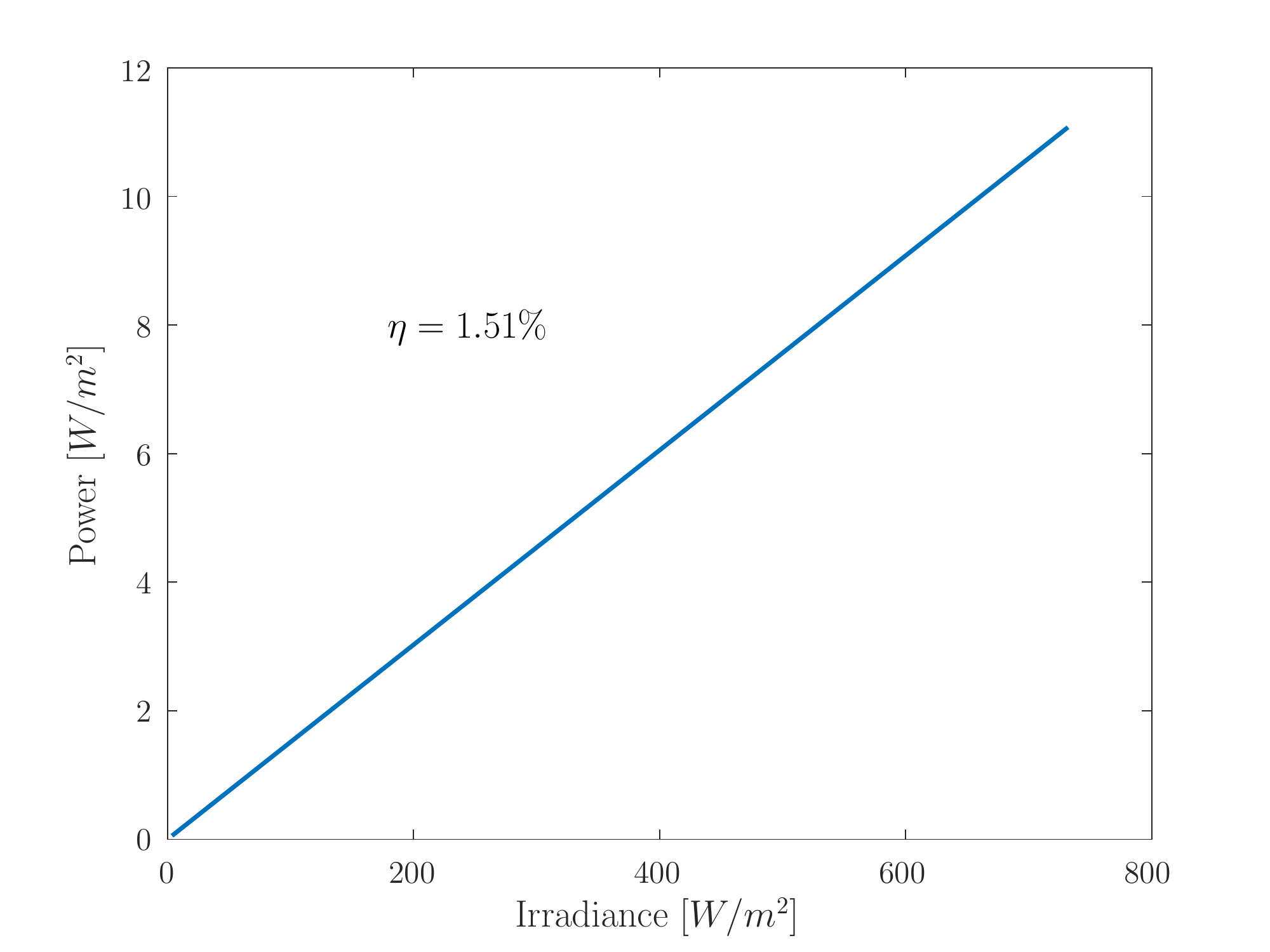}\caption{\label{fig:Device-output-power}Device output power as a function
of the irradiance}
\end{figure}

After depositing a 150nm ITO coating and adding the bus bar, fingers
and back electrode, the photo-current was measured under various intensities
which were controlled using a UNI-T UT383BT luxmeter. The measured
illuminance was converted to irradiance based on a recent guide which
tries to sort out the various conversion factors \cite{michael2020conversion}.
Based on their recommendation we used a $116\,\frac{lux}{W/m^{2}}$.
Figure \ref{fig:Device-output-power} displays the devices output
power as a function of the irradiance.

The slope of the graph shows the efficiency of the device, which is
1.51\% in the above measurement. 

\section{Conclusions}

This work presents a viable method for recycling and re-purposing
used silicon wafers for PV use. The preliminary work shows that standard
stripping methods followed by a newly suggested method for diffusing
dopants does provide a working PV cell. Both optical and electrical
properties of the resulting device were presented. The output power
density's dependence on the incident irradiance resulted in a $1.51\,\%$
efficiency, which is an order of magnitude lower than expected. As
this work was intended as a feasibility study, the result shows promise.
Following this work we intend to improve the efficiency of a single
crystal wafer cell and extend the work for recycling polycrystalline
wafers which are abundant in the ever increasing retired solar panels.

\bibliographystyle{unsrt}
\bibliography{references_PV}

\end{document}